\documentclass[aps,prl,showpacs,superscriptaddress,twocolumn,longbibliography]{revtex4-1}

\usepackage{amsfonts}
\usepackage{subfigure}
\usepackage{amsmath}
\usepackage{amssymb}
\usepackage{amsbsy} 
\usepackage{epsfig}
\usepackage{graphicx}
\usepackage{epstopdf}

\def\be{\begin{equation}} \def\ee{\end{equation}}
\def\bea{\begin{eqnarray}} \def\eea{\end{eqnarray}}

\def\nn{\nonumber}

\def\be{{\bf e}}

\newcommand{\bra}[1]{\langle#1|}
\newcommand{\ket}[1]{|#1\rangle}

\def\rw{\rightarrow}

\def\la{\langle}
\def\ra{\rangle}

\begin{document}

\title{Non-Hermitian skin effect and chiral damping in open quantum systems}

\author{Fei Song}
 \affiliation{ Institute for
Advanced Study, Tsinghua University, Beijing,  100084, China }

\author{Shunyu Yao}
 \affiliation{ Institute for
Advanced Study, Tsinghua University, Beijing,  100084, China }

\author{Zhong Wang} \altaffiliation{ wangzhongemail@gmail.com }
\affiliation{ Institute for
Advanced Study, Tsinghua University, Beijing,  100084, China }


\begin{abstract}

One of the unique features of non-Hermitian Hamiltonians is the non-Hermitian skin effect, namely that the eigenstates are exponentially localized at the boundary of the system. For open quantum systems, a short-time evolution can often be well described by the effective non-Hermitian Hamiltonians, while long-time dynamics calls for the Lindblad master equations, in which the Liouvillian superoperators generate time evolution. In this Letter, we find that Liouvillian superoperators can exhibit the non-Hermitian skin effect, and uncover its unexpected physical consequences. It is shown that the non-Hermitian skin effect dramatically shapes the long-time dynamics, such that the damping in a class of open quantum systems is algebraic under periodic boundary condition but exponential under open boundary condition. Moreover, the non-Hermitian skin effect and non-Bloch bands cause a chiral damping with a sharp wavefront. These phenomena are beyond the effective non-Hermitian Hamiltonians; instead, they belong to the non-Hermitian physics of full-fledged open quantum dynamics.

\end{abstract}

\maketitle

Non-Hermitian Hamiltonians provide a natural framework for a wide range of phenomena such as photonic systems with loss and gain\cite{feng2017non,el2018non,ozawa2018rmp,peng2014lossinduced,feng2013experimental}, open quantum systems\cite{rotter2009non,zhen2015spawning, diehl2011topology,verstraete2009quantum,malzard2015open,Dalibard1992, carmichael1993,Anglin1997,choi2010coalescence,diehl2008quantum,bardyn2013topology}, and quasiparticles with finite lifetimes\cite{kozii2017,papa2018bulk,shen2018quantum,Zhou2018arc,Yoshida2018heavy}. Recently, the interplay of non-Hermiticity and topological phases have been attracting growing attentions. Considerable attentions have been focused on non-Hermitian bulk-boundary correspondence\cite{shen2017topological, lee2016anomalous,yao2018edge,yao2018chern,Alvarez2018,leykam2017, kunst2018biorthogonal,xiong2017,alvarez2017, Yokomizo2019,jin2019,Zirnstein2019,herviou2018restoring}, new topological invariants\cite{yao2018edge,yao2018chern,esaki2011,leykam2017,gong2018nonhermitian, liu2019second,lieu2018ssh,yin2018ssh,Yokomizo2019,Deng2019, Ghatak2019,jiang2018invariant}, generalizations of topological insulators\cite{harari2018topological,yuce2015topological,yao2018chern, zhu2014PT,lieu2018bdg,yuce2016majorana,menke2017, wang2019nonQuantized,Philip2018loss,chen2018hall,hirsbrunner2019topology, klett2017sshkitaev,Zeng2019quasicrystal,zhou2018floquet,kawabata2018PT} and semimetals\cite{xu2017weyl,wang2019nodal,cerjan2018weyl,budich2019,yang2019hopf,Carlstrom2018, Okugawa2019,Moors2019,zyuzin2018flat,yoshida2018exceptional,zhou2018exceptional}, and novel topological classifications\cite{kawabata2019unification,zhou2019periodic,kawabata2018symmetry}, among other interesting theoretical\cite{rudner2009topological,McDonald2018phase, hu2017exceptional, Silveirinha2019,kawabata2019exceptional,Louren2018kondo, rui2019,Bliokh2019,Luo2019higher,Turker2019open,Hatano1996} and experimental\cite{zeuner2015bulk,xiao2017observation,poli2015selective, weimann2017topologically,cerjan2018experimental,zhan2017detecting} investigations.

One of the remarkable phenomena of non-Hermitian systems is the \emph{non-Hermitian skin effect}\cite{yao2018edge,Alvarez2018}(NHSE), namely that the majority of eigenstates of a non-Hermitian operator are localized at boundaries, which suggests the non-Bloch bulk-boundary correspondence\cite{yao2018edge,kunst2018biorthogonal} and non-Bloch band theory based on the generalized Brillouin zone\cite{yao2018edge,yao2018chern,liu2019second,Yokomizo2019,Deng2019}. Broader implications of NHSE have been under investigations\cite{yao2018chern,jiang2019interplay,lee2018anatomy,lee2018hybrid, jin2019,Zirnstein2019,kunst2018transfer,liu2019second,Edvardsson2019,Borgnia2019, wang2019nodal,wang2018phonon,ezawa2018higher,Ezawa2018,ezawa2018electric,yang2019non,ge2019topological}.
Very recently, NHSE has been observed in experiments\cite{Helbig2019NHSE,Xiao2019NHSE,Ghatak2019NHSE}.

In open quantum systems, non-Hermiticity naturally arises in the Lindblad master equation that governs the time evolution of density matrix (see e.g., Refs.\cite{diehl2011topology,verstraete2009quantum}): \bea
\frac{d\rho}{dt}= -i[H,\rho]+\sum_{\mu}\left(2L_{\mu}\rho L_\mu^{\dag}-\{L_\mu^\dag L_\mu,\rho\}\right) \equiv   \mathcal{L}\rho, \label{master}
\eea where $H$ is the Hamiltonian, $L_\mu$'s are the Lindblad dissipators describing quantum jumps due to coupling to the environment, and $\mathcal{L}$ is called the Liouvillian superoperator. Before the occurrence of a jump, the short-time evolution follows the effective non-Hermitian Hamiltonian $H_\text{eff}=H -i\sum_\mu L^\dag_\mu L_\mu$ as $d\rho/dt = -i(H_\text{eff}\rho -\rho H^\dag_\text{eff})$\cite{Dalibard1992,carmichael1993,Nakagawa2018kondo}.

It is generally believed that when the system size is not too small,     the effect of boundary condition is insignificant. As such, periodic boundary condition is commonly adopted, though open-boundary condition is more relevant to experiments. In this paper, we show that the long time Lindblad dynamics of an open-boundary system dramatically differ from that of a periodic-boundary system. Furthermore, this is related to the NHSE of the damping matrix derived from the Liouvillian. Notable examples are found that the long time damping is algebraic (i.e., power law) under periodic boundary condition while exponential under open boundary condition. Moreover, NHSE implies that the damping is unidirectional, which is dubbed the ``chiral damping''. Crucially, the theory is based on the full Liouvillian. Although $H_\text{eff}$ may be expected to play an important role, it is in fact inessential here (e.g., its having NHSE or not does not matter).

\begin{figure}
\includegraphics[width=7cm, height=3.7cm]{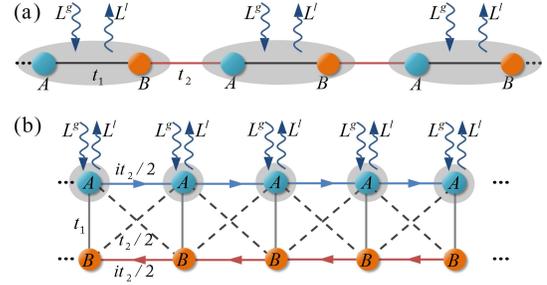}
\caption{ (a) SSH model with staggered hopping $t_1$ and $t_2$, with the ovals indicating the unit cells. The Bloch Hamiltonian is $h(k)=(t_1+t_2\cos k)\sigma_x +t_2\sin k\sigma_y$. The fermion loss and gain are described by the dissipators $L^l$ and $L^g$ [Eq.(\ref{dissipator})] in the master equation framework.  (b) A different realization of the same model. The hopping Hamiltonian $h(k)=(t_1+t_2\cos k)\sigma_x +t_2\sin k\sigma_z$, and the dissipators are $L^l_x=\sqrt{\gamma_l}c_{xA}$ and $L^g_x=\sqrt{\gamma_g}c^\dag_{xA}$. (b) is equivalent to (a) via a basis change $\sigma_y\leftrightarrow  \sigma_z$. Because the gain and loss is on-site, (b) is more feasible experimentally.     } \label{illustration}
\end{figure}

\emph{Model.--}The system is illustrated in Fig.\ref{illustration}(a). Our Hamiltonian $H=\sum_{ij}h_{ij}c^{\dag}_{i}c_{j}$, where $c^\dag_i,c_i$ are fermion creation and annihilation operators at site $i$ (including additional degrees of freedom such as spin is straightforward). We will consider single particle loss and gain, with loss dissipators $L_\mu^l=\sum_i D_{\mu i}^lc_i$ and gain dissipators $L_\mu^g=\sum_i D_{\mu i}^gc_i^\dag$, respectively. For concreteness, we take $h$ to be the Su-Schrieffer-Heeger (SSH) Hamiltonian, namely $h_{ij}=t_1$ and $t_2$ on adjacent links. A site is also labelled as $i=xs$, where $x$ refers to the unit cell, and $s=A,B$ refers to the sublattice. For simplicity, let each unit cell contain a single loss and gain dissipator (namely, $\mu$ is just $x$): \bea L_x^l=\sqrt{ \gamma_l/2}(c_{xA}-ic_{xB}),\quad L_x^g=\sqrt{ \gamma_g/2}(c_{xA}^\dag+ic_{xB}^\dag); \nn\\ \label{dissipator} \eea  in other words, $D^l_{x,xA}=iD^l_{x,xB}=\sqrt{\gamma_l/2}, D^g_{x,xA}=-iD^g_{x,xB}=\sqrt{\gamma_g/2}$. We recognized in Eq.(\ref{dissipator}) that the $\sigma_y=+1$ states are lost to or gained from the bath. A seemingly different but essentially equivalent realization of the same model is shown in Fig.\ref{illustration}(b), which can be obtained from the initial model [Fig.\ref{illustration}(a)] after a basis change $\sigma_y\leftrightarrow\sigma_z$. Accordingly, the dissipators in Fig.\ref{illustration}(b) are $\sigma_z=1$ states. As the gain and loss are on-site, its experimental implementation is easier. Keeping in mind that Fig.\ref{illustration}(b) shares the same physics, hereafter we focus on the setup in Fig.\ref{illustration}(a).

To see the evolution of density matrix, it is convenient to monitor the single-particle correlation $\Delta_{ij}(t)=\text{Tr}[c^{\dag}_{i}c_{j}\rho(t)]$, whose time evolution is $d\Delta_{ij}/dt=\text{Tr}[c^{\dag}_{i}c_{j}d\rho/dt]$. It follows from Eq.(\ref{master}) that (see the Supplemental Material)
\bea
\frac{d \Delta(t)}{dt}=i[h^T,\Delta(t)]-\{M^T_l+M_g,\Delta(t)\}+2M_g, \label{evolution-1}
\eea
where $(M_g)_{ij}=\sum_\mu D_{\mu i}^{g*}D_{\mu j}^g$ and $(M_l)_{ij}=\sum_\mu D_{\mu i}^{l*}D_{\mu j}^l$, and both $M_l$ and $M_g$ are Hermitian matrices. Majorana versions of Eq.(\ref{evolution-1}) appeared in Refs.\cite{diehl2011topology,Prosen2011,bardyn2013topology}. We can define the damping matrix \bea X=ih^T-(M_l^T+M_g), \label{damping} \eea which recasts Eq.(\ref{evolution-1}) as \bea \frac{d \Delta(t)}{dt}= X\Delta(t) +\Delta(t) X^\dag +2M_g. \label{evolution2} \eea
The steady state correlation $\Delta_s=\Delta(\infty)$, to which long time evolution of any initial state converges, is determined by $d\Delta_s/dt=0$, or $
X\Delta_s+\Delta_sX^\dag+2M_g=0$.
In this paper, we are concerned mainly about the dynamics, especially the speed of converging to the steady state, therefore, we shall focus on the deviation  $\tilde{\Delta}(t)=\Delta(t)-\Delta_s$, whose evolution is $d \tilde{\Delta}(t)/dt= X\tilde{\Delta}(t) +\tilde{\Delta}(t) X^\dag$, which is readily integrated to
\bea
\tilde{\Delta}(t)=e^{Xt}\tilde{\Delta}(0)e^{X^\dag t}.\label{evolve}
\eea
We can write $X$ in terms of right and left eigenvectors\footnote{As a non-Hermitian matrix, $X$ can have exceptional points, and we have checked that our main results are qualitatively similar therein. }, \bea X= \sum_n \lambda_n |u_{Rn}\ra\la u_{Ln}|, \label{Xexpansion} \eea and express Eq.(\ref{evolve}) as \bea \tilde{\Delta}(t)=\sum_{n,n'}\exp[(\lambda_n+\lambda_{n'}^*)t]|u_{Rn}\ra\la u_{Ln}|\tilde{\Delta}(0)|u_{Ln'}\ra\la u_{Rn'}|.\nn\\ \label{expansion} \eea  By the dissipative nature,  $\text{Re}(\lambda_n)\leq 0$ always holds true. The Liouvillian gap $\Lambda=\text{min}[2\text{Re}(-\lambda_n)]$ is decisive for the long-time dynamics. A finite gap implies exponential converging rate towards the steady state, while a vanishing gap implies algebraic convergence\cite{Cai2013algebraic}.

\begin{figure}
\includegraphics[width=7cm, height=4.6cm]{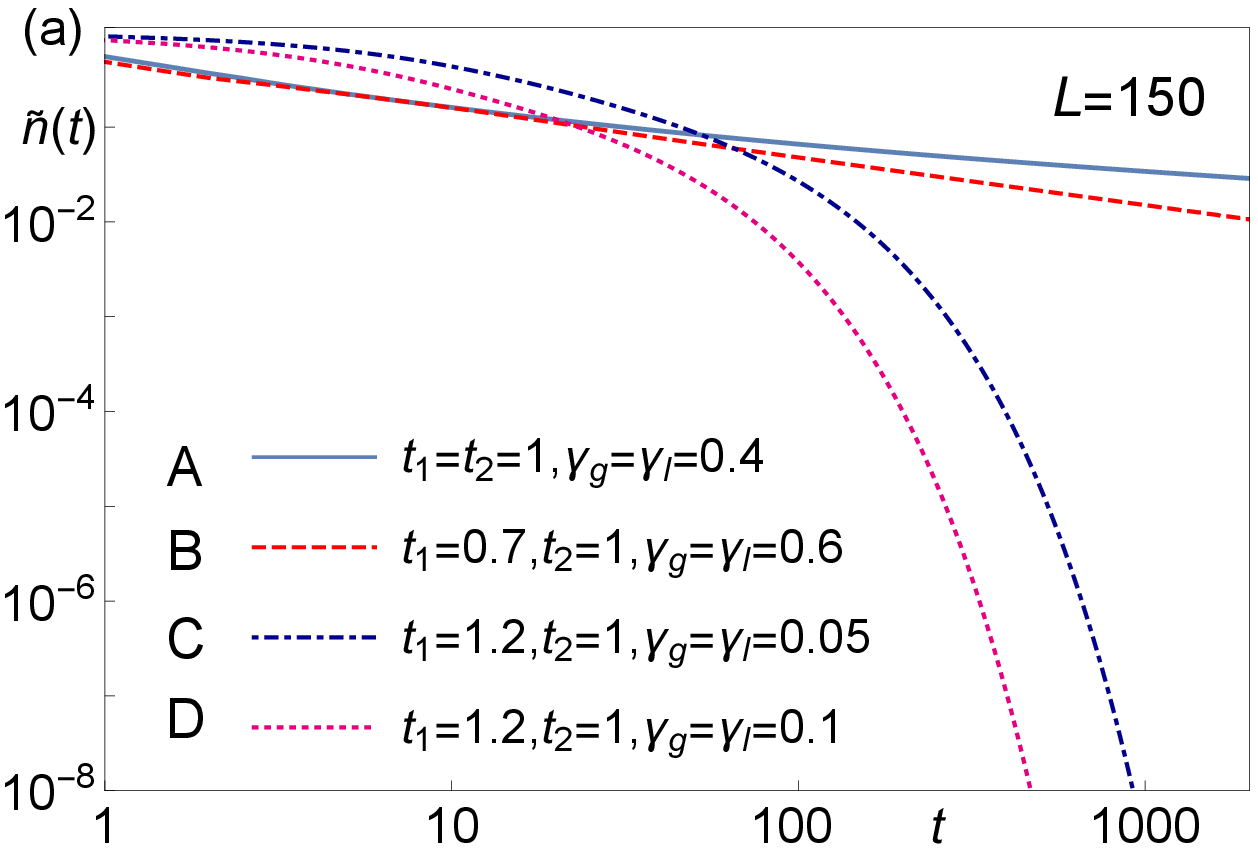}
\includegraphics[width=7cm, height=6.3cm]{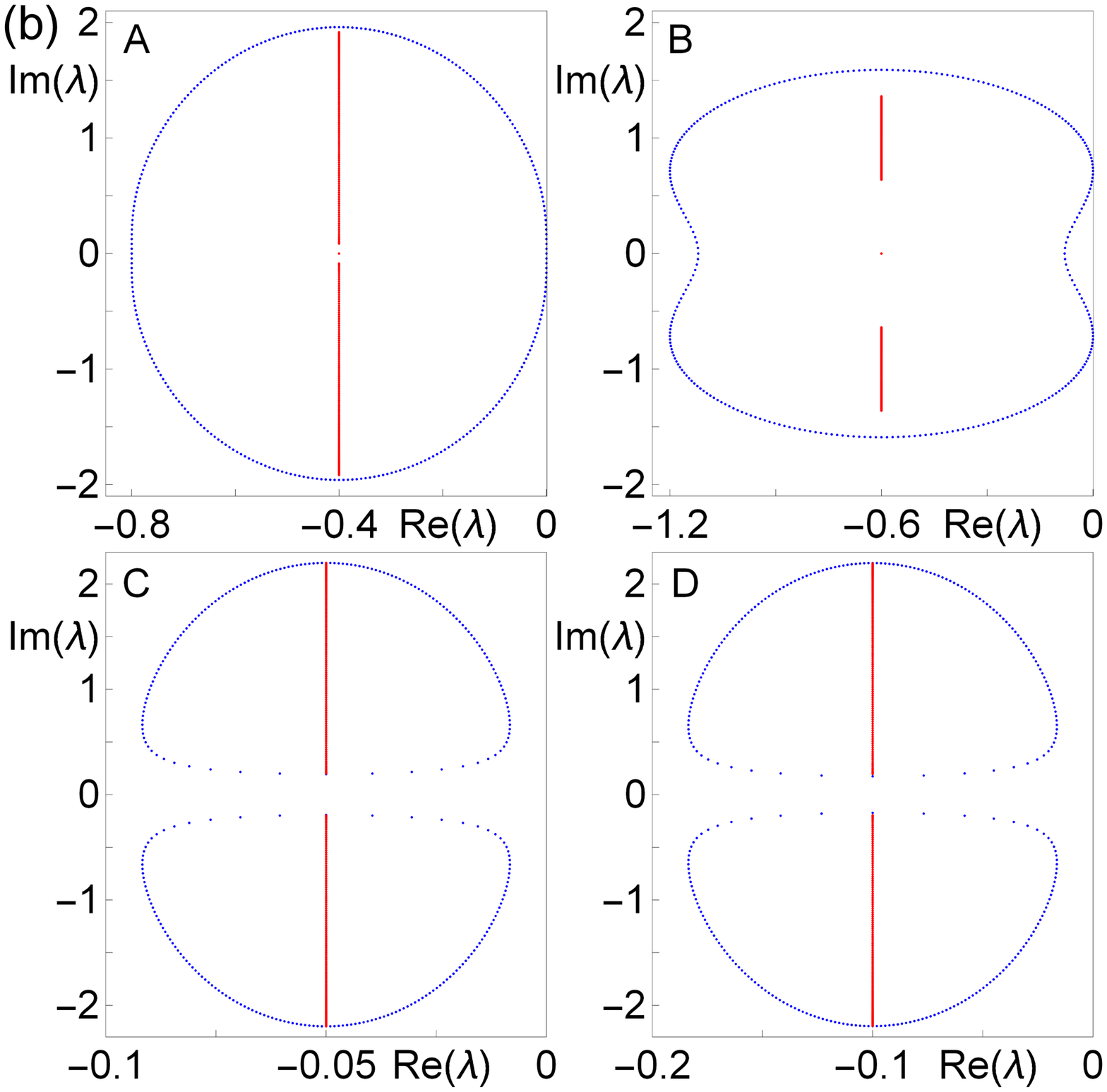}
\caption{(a) The damping of fermion number towards the steady state of a periodic boundary chain with length $L=150$ (unit cell). The damping is algebraic for cases $A,B$ with $t_1\leq t_2$, while exponential for $C,D$ with $t_1>t_2$. The initial state is the completely filled state $\prod_{x,s}c^\dag_{xs}|0\ra$. (b) The eigenvalues of the damping matrix $X$. Blue: periodic-boundary; Red:  open-boundary. The Liouvillian gap of the periodic-boundary chain vanishes for $A$ and $B$, while it is nonzero for $C$ and $D$. For the open-boundary chain, the Liouvillian gap is nonzero in all four cases. This drastic spectral distinction between open and periodic boundary comes from the NHSE (see text).  } \label{fig2}
\end{figure}

\begin{figure}
\includegraphics[width=7.0cm, height=4.3cm]{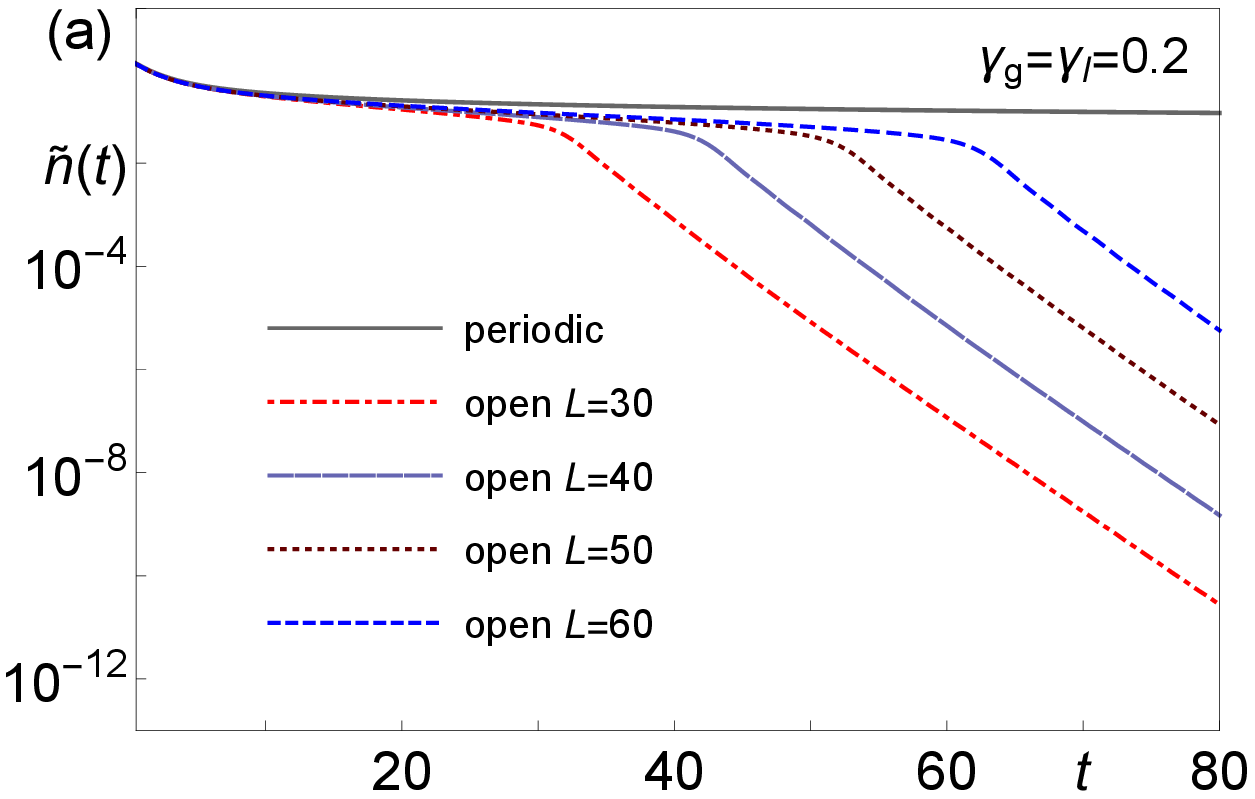}
\includegraphics[width=7.0cm, height=4.3cm]{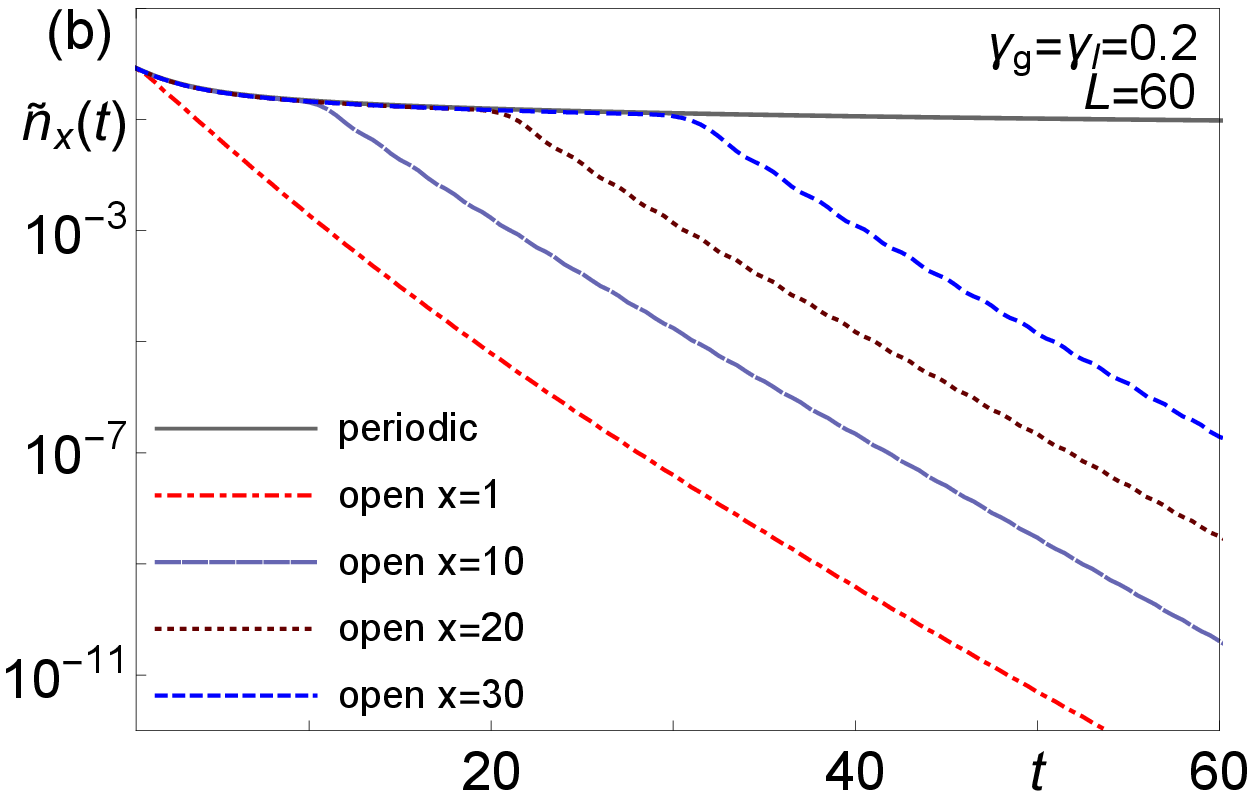}
\caption{(a) The particle number damping of a periodic boundary chain (solid curve) and open-boundary chains for several chain length $L$. The long time damping of a periodic chain follows a power law, while the open boundary chain follows an exponential law after an initial power law stage.  (b) The site-resolved damping.  The left end ($x=1$) enters the exponential stage from the very beginning, followed sequentially by other sites. For both (a) and (b), the initial state is the completely filled state $\prod_{x,s}c^\dag_{xs}|0\ra$, therefore, $\Delta(0)$ the identity matrix $I_{2L\times 2L}$.   $t_1=t_2=1,\gamma_g=\gamma_l=0.2$.   }\label{fig3}
\end{figure}

\emph{Periodic chain.--}Let us study the periodic boundary chain, for which going to momentum space is more convenient. It can be readily found that $h(k)=(t_1+t_2\cos k)\sigma_x+t_2\sin k\sigma_y$ and \bea M_l(k)=\frac{\gamma_l}{2}(1+\sigma_y), \quad M_g(k)=\frac{\gamma_g}{2}(1-\sigma_y).\eea These $M(k)$ matrices are $k$-independent because the gain and loss dissipators are intra-cell. The Fourier transformation of $X$ is $X(k)=ih^T(-k)-M_l^T(-k)-M_g(k)$ (the minus sign in $-k$ comes from matrix transposition), therefore, the damping matrix in momentum space reads
\bea
X(k) = i[(t_1+t_2\cos k)\sigma_x+ (t_2\sin k-i\frac{\gamma}{2})\sigma_y] -\frac{\gamma}{2}I, \nn\\ \label{Xexpression}
\eea  where $\gamma\equiv\gamma_l+\gamma_g$. If we take the realization in Fig.\ref{illustration}(b) instead of Fig.\ref{illustration}(a), the only modification to $X(k)$ is a basis change $\sigma_y\rw\sigma_z$ in Eq.(\ref{Xexpression}), with the physics unchanged.  Diagonalizing $X(k)$, we find that the Liouvillian gap $\Lambda=0$ for $t_1\leq t_2$, while the gap opens for $t_1>t_2$ [Fig.\ref{fig2}(b)]. The damping rate is therefore expected to be algebraic and exponential in each case, respectively. To confirm this, we numerically calculate the site-averaged fermion number deviation from the steady state, defined as $\tilde{n}(t)=\sqrt{\sum_x \tilde{n}^2_x(t)/L}$, where     $\tilde{n}_x(t)=n_x(t)-n_x(\infty)$ with $n_x(t)=n_{xA}(t)+n_{xB}(t)$, $n_{xs}\equiv\Delta_{xs,xs}$ being the fermion number at site $xs$. The results are consistent with the vanishing (nonzero) gap in the $t_1\leq t_2$ ($t_1> t_2$) case [Fig.\ref{fig2}(a)].

Although our focus here is the damping dynamics, we also give the steady state. In fact, our $M_l$ and $M_g$ satisfy $M_l^T+M_g=M_g \gamma/\gamma_g $, which guarantees that $\Delta_s=(\gamma_g/\gamma)I_{2L\times 2L}$ is the steady state solution. It is independent of boundary conditions.

Now we show the direct relation between the algebraic damping and the vanishing gap of $X$. The eigenvalues of $X(k)$ are \bea \lambda_{\pm}(k)=-\gamma/2\pm i\sqrt{(t_1^2+t_2^2+2t_1t_2\cos k-\gamma^2/4)-it_2\gamma\sin k}. \quad\quad \label{lambda} \eea  Let us consider $t_1=t_2\equiv t_0$ for concreteness (case $A$ in Fig.\ref{fig2}), then $\lambda_{-}(\pi)=0$ and the expansion in $\delta k\equiv k-\pi$ reads
\bea
\lambda_{-}(\pi+\delta k)\approx -it_0\delta k-\frac{t_0^2}{4\gamma}(\delta k)^4.
\eea
Now Eq.(\ref{Xexpansion}) becomes $X=\sum_{k,\alpha=\pm}\lambda_{\alpha}(k)\ket{u_{Rk\alpha}}\bra{u_{Lk\alpha}}$, and Eq.(\ref{expansion}) reads
\bea
\tilde{\Delta}(t)= \sum_{kk',\alpha\alpha'}e^{\lambda_\alpha(k)t+\lambda_{\alpha'}^*(k')t} \ket{u_{Rk\alpha}}\bra{u_{Lk\alpha}}\tilde{\Delta}(0) \ket{u_{Lk'\alpha'}}\bra{u_{Rk'\alpha'}}. \nn \\ 
\eea
For the initial state with translational symmetry, we have $\bra{u_{Lk\alpha}}\tilde{\Delta}(0)\ket{u_{Lk'\alpha'}} =\delta_{kk'}\bra{u_{Lk\alpha}}\tilde{\Delta}(0)|u_{Lk\alpha'}\rangle$. The long-time behavior of $\tilde{\Delta}(t)$ is dominated by the $\alpha=\alpha'=-$ sector, which provides a decay factor $\sum_{\delta k} \exp\left(2\text{Re}[\lambda_-(\pi+\delta  k)]t\right) \approx \int d(\delta k)\exp[- \frac{t_0^2}{2\gamma} (\delta k)^4t]\sim t^{-1/4}$. Similarly, for $t_1<t_2$ we have $\tilde{\Delta}(t)\sim t^{-1/2}$.

\emph{Chiral damping.--}Now we turn to the open boundary chain. Although the physical interpretation is quite different, our $X$ matrix resembles the non-Hermitian SSH Hamiltonian\cite{yao2018edge,yin2018ssh}, as can be appreciated from Eq.(\ref{Xexpression}). Remarkably, all the eigenstates of $X$ are exponentially localized at the boundary (i.e., NHSE\cite{yao2018edge}). As such, the eigenvalues of open boundary $X$ cannot be obtained from $X(k)$ with real-valued $k$; instead, we have to take complex-valued wavevectors $k+i\kappa$. In other words, the usual Bloch phase factor $e^{ik}$ living in the unit circle is replaced by $\exp[i(k+i\kappa)]$ inhabiting a generalized Brillouin zone\cite{yao2018edge}, whose shape can be precisely calculated in the non-Bloch band theory\cite{yao2018edge,yao2018chern,Yokomizo2019,liu2019second,Deng2019}.

From the non-Bloch band theory\cite{yao2018edge}, we find that
$\kappa=-\ln \sqrt{\left|\frac{t_1+\gamma/2}{t_1-\gamma/2}\right|}$,
and that the eigenvalues of $X$ of an open boundary chain are
$\lambda_{\pm}(k+i\kappa)$, where $\lambda_\pm$ are the $X(k)$ eigenvalues given in Eq.(\ref{lambda}). We can readily check that, for $|\gamma|<2|t_1|$,  \bea \lambda_{\pm}(k+i\kappa)= -\frac{\gamma}{2} \pm i E(k), \label{openlambda} \eea where $E(k)=\sqrt{t_1^2+t_2^2-\frac{\gamma^2}{4}+2 t_2\sqrt{t_1^2-\frac{\gamma^2}{4}}\cos k}$, which is real. We have also numerically diagonalized $X$ for a long open chain [red dots in Fig.\ref{fig2}(b)], which confirms Eq.(\ref{openlambda}). An immediate feature of Eq.(\ref{openlambda}) is that the real part is a constant, $-\gamma/2$, which is consistent with the numerical spectrums [Fig.\ref{fig2}(b))]. We note that the analytic results  based on generalized Brillouin zone produce the continuum bands only, and the isolated topological edge modes [Fig.\ref{fig2}(b), A and B panels] are not contained in Eq.(\ref{openlambda}), though they can be inferred from the non-Bloch bulk-boundary correspondence\cite{yao2018edge,kunst2018biorthogonal}. Here, we focus on bulk dynamics, and these topological edge modes do not play important roles\footnote{Topological edge modes have been investigated recently in Ref. \cite{Caspel2019,Kastoryano2019} in models without NHSE.}.

It follows from Eq.(\ref{openlambda}) that the Liouvillian gap $\Lambda=\gamma$, therefore, we expect an exponential long-time damping of $\tilde{\Delta}(t)$. This exponential behavior has been confirmed by numerical simulation [Fig.\ref{fig3}(a)]. Before entering the exponential stage, there is an initial period of algebraic damping, whose duration grows with chain length $L$ [Fig.\ref{fig3}(a)]. To better understand this feature, we plot the damping in each unit cell [Fig.\ref{fig3}(b)]. We find that the left end ($x=1$) enters the exponential damping immediately, and other sites enter the exponential stage sequentially, according to their distances to the left end. As such, there is a ``damping wavefront'' traveling from the left (``upper reach'') to right (``lower reach''). This is dubbed a ``chiral damping'', which can be intuitively related to the fact that all eigenstates of $X$ are localized at the right end\cite{yao2018edge}.

\begin{figure}
\includegraphics[width=9.6cm, height=4cm]{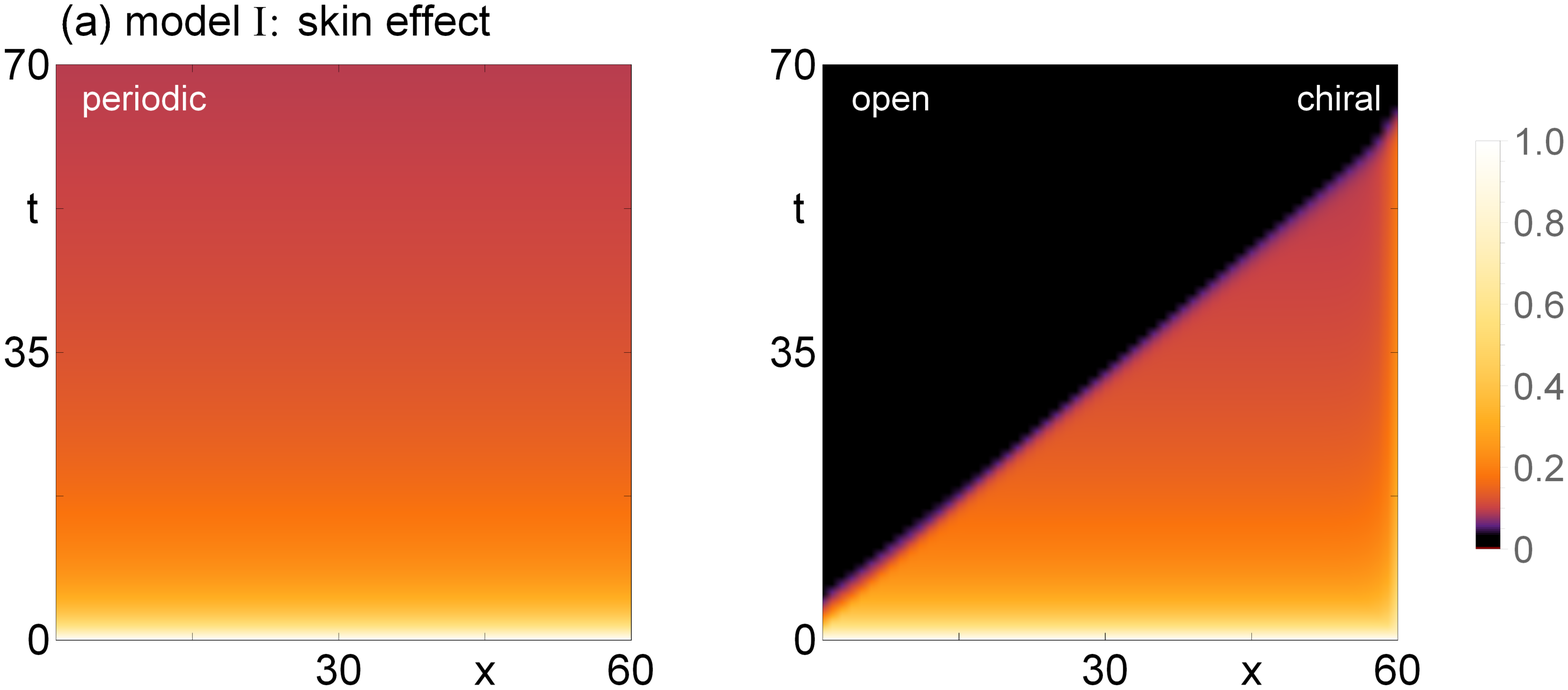}
\includegraphics[width=9.6cm, height=4cm]{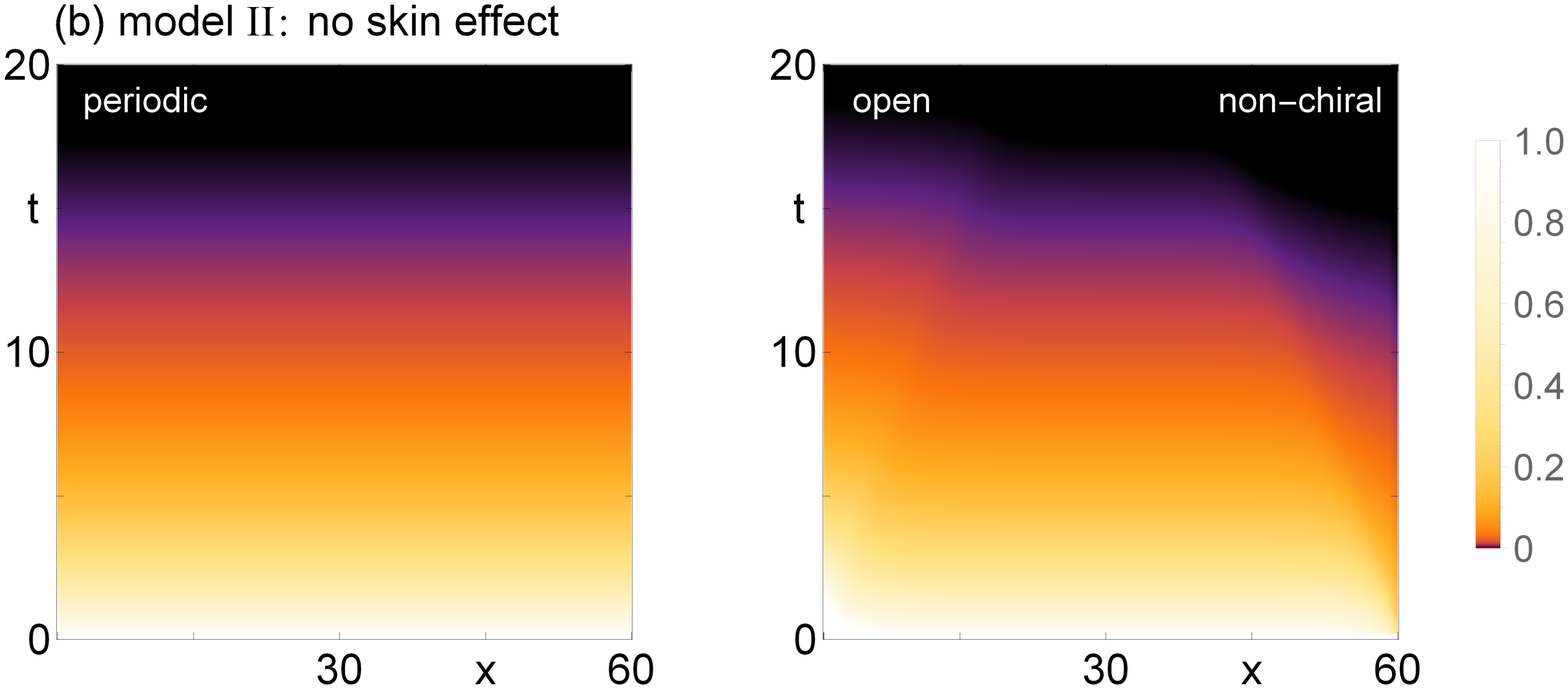}
\caption{ Time evolution of $\tilde{n}_x(t)=n_x(t)-n_x(\infty)$, which shows damping of particle number $n_x(t)$ towards the steady state. (a) $\tilde{n}_x(t)$ of the main model  with dissipators given by Eq.(\ref{dissipator}) (referred to as ``model I''). Left: periodic boundary; Right: open boundary. The chiral damping is clearly seen in the open boundary case. The dark region corresponds to the exponential damping stage seen in Fig.\ref{fig3}.  (b) $\tilde{n}_x(t)$ of model II, whose damping matrix $X$ [Eq.(\ref{XII})] has no NHSE. The Liouvillian gap is nonzero and the same for periodic and open boundary chains. Common parameters: $t_1=t_2=1; \gamma_g=\gamma_l=0.2$. \label{chiral} }
\end{figure}

More intuitively, the damping of $\tilde{n}_x(t)= n_x(t)-n_x(\infty)$ is shown in Fig.\ref{chiral}(a). In the periodic boundary chain it follows a slow power law. In the open boundary chain, a right-moving wavefront is seen. After the wavefront passes by $x$, the algebraically decaying $\tilde{n}_x(t)$ enters the exponential decay stage and rapidly diminishes.

The wavefront can be understood as follows. According to Eq.(\ref{evolve}), the damping of $\tilde{\Delta}(t)$ is determined by the evolution under $\exp(Xt)$, which is just the evolution under $\exp(-\gamma t/2)\exp(-iH_\text{SSH}t)$, where $H_\text{SSH}$ is the non-Hermitian SSH Hamiltonian\cite{yao2018edge} (with an unimportant sign difference). Now the propagator $\la xs|\exp(-iH_\text{SSH}t)|x's'\ra$  can be decomposed as propagation of various momentum modes with velocity $v_k=\partial E/\partial k$. Due to the presence of an imaginary part $\kappa$ in the momentum, propagation from $x'$ to $x$ acquires an $\exp[-\kappa(x-x')]$ factor. If this factor can compensate $\exp(-\gamma t/2)$, exponential damping can be evaded, giving way to a power law damping. For simplicity we take $\gamma$ small, so that $\kappa\approx -\gamma/2t_1$, therefore $\exp[-\kappa(x-x')]\approx \exp[v_k(\gamma/2t_1)t]$ and the damping of propagation from $x'$ to $x$ is $\exp[(-\gamma/2+v_k\gamma/2t_1)t]$ for the $k$ mode. By a straightforward calculation, we have $\max(v_k)=t_2$ (for $t_1>t_2$) or $\sqrt{|t_1^2-\gamma^2/4|}\approx t_1$ (for $t_1\leq t_2$). Let us consider $t_1\leq t_2$ first. When $x>\max(v_k)t$, the propagation from $x'=x-\max(v_k)t$ to $x$ carries a factor $\exp[(-\gamma/2+\max(v_k)\gamma/2t_1)t]=1$; while for $x<\max(v_k)t$, we need the nonexistent $x'=x-\max(v_k)t<0$, therefore, compensation is impossible and we have exponential damping. This indicates a wavefront at $x=\max(v_k)t$. For $t_1=t_2=1$, we have $\max(v_k)\approx 1$, which is consistent with the wavefront velocity ($\approx 1$) in Fig.\ref{chiral}(a).

As a comparison, we introduce the ``model II'' (the model studied so far is referred to as ``model I'') that differs from model I only in $L_x^g$, which is now $L_x^g=\sqrt{\frac{\gamma_g}{2}}(c_{xA}^\dag-ic_{xB}^\dag)$ [compare it with Eq.(\ref{dissipator})]. The damping matrix is \bea
X(k) = i[(t_1+t_2\cos k)\sigma_x+ (t_2\sin k-i\frac{\gamma_l-\gamma_g}{2})\sigma_y] -\frac{\gamma}{2}I, \quad \quad \label{XII} \eea  which has no NHSE when $\gamma_l=\gamma_g$. Accordingly, the open and periodic boundary chains have the same Liouvillian gap, and chiral damping is absent [Fig.\ref{chiral}(b)].

In realistic systems, there may be disorders, fluctuations of parameters, and other imperfections. Fortunately, the main results here are based on the presence of NHSE, which is a quite robust phenomenon unchanged by modest imperfections. As such, it is expected that our main predictions are robust and observable.

\emph{Final remarks.--}(i) The chiral damping originates from the NHSE of the damping matrix $X$ rather than the effective non-Hermitian Hamiltonian. Unlike the damping matrix, the effective non-Hermitian Hamiltonian describes short time evolution. It is found to be
$H_\text{eff} = \sum_{ij}c^\dag_i (h_\text{eff})_{ij} c_j -i \gamma_g L$, where $h_\text{eff}$, written in momentum space, is
$h_\text{eff}(k) =(t_1+t_2\cos k)\sigma_x+(t_2\sin k-i\frac{\gamma_l-\gamma_g}{2})\sigma_y  -i\frac{\gamma_l-\gamma_g}{2}I$. For $\gamma_g=\gamma_l$, $h_\text{eff}$ has no NHSE, though $X$ has. Although damping matrices with NHSE can arise quite naturally (e.g., in Fig.\ref{illustration}), none of the previous models (e.g., Ref.\cite{Ashida2018}) we have checked has NHSE.

(ii) The periodic-open contrast between the slow algebraic and fast exponential damping has important implications for experimental preparation of steady states (e.g. in cold atom systems). In the presence of NHSE, approaching the steady states in open-boundary systems can be much faster than estimations based on periodic boundary condition.

(iii) It is interesting to investigate other rich aspects of non-Hermitian physics such as PT symmetry breaking\cite{peng2014parity} in this platform (Here, we have focused on the cases that the open-boundary $iX$ is essentially PT-symmetric, meaning that the real parts of $X$ eigenvalues are constant).

(iv) When fermion-fermion interactions are included, higher-order correlation functions are coupled to the two-point ones, and approximations (such as truncations) are called for. Moreover, the steady states can be multiple\cite{Albert2018,Zhou2017dissipative}, in which case the damping matrix depends on the steady state approached, leading to even richer chiral damping behaviors. These possibilities will be left for future investigations.

This work is supported by NSFC under grant No. 11674189.

\bibliography{dirac}

\vspace{11mm}

\section{Derivation of the differential equation of correlation functions}

We shall derive Eq.(3) in the main article, which is reproduced as follows:
\bea
\frac{d \Delta(t)}{dt}=i[h^T,\Delta(t)]-\{M^T_l+M_g,\Delta(t)\}+2M_g. \label{supp} \eea

In fact, after inserting the Lindblad master equation into $d\Delta_{ij}/dt=\text{Tr}[c^{\dag}_{i}c_{j}d\rho/dt]$ and reorganize the terms, we have
\bea
\frac{d\Delta_{ij}(t)}{dt} &=& i\text{Tr}\left([H,c^\dagger_i c_j] \rho(t) \right)+ \nn \\ && \sum_\mu \text{Tr}\left[\left(2L^\dagger_\mu[c^\dagger_i c_j,L_\mu]+[L^\dagger_\mu L_\mu,c^\dagger_i c_j]\right)\rho(t) \right].
  \label{EOM}
\eea
By a straightforward calculation, we have
\bea
 [H,c_i^\dagger c_j]  &=&\sum_{mn}h_{mn}[c^\dagger_m c_n,c_i^\dagger c_j] \nn \\
&=&\sum_{mn}h_{mn}(-\delta_{mj}c_i^\dagger c_n+\delta_{in}c_m^\dagger c_j) \nn \\
&=&\sum_n (-h_{jn} c_i^\dagger c_n + h_{ni}c_n^\dagger c_j),
\eea
therefore, the Hamiltonian commutator term in Eq.(\ref{EOM}) is reduced to $i[h^T,\Delta(t)]_{ij}$, which is the first term of Eq.(\ref{supp}).

The commutators terms from the loss dissipators $L_\mu^l=\sum_i D_{\mu i}^lc_i$ are
\bea
 2\sum_\mu L_\mu^{l \dagger}[c^\dagger_i c_j,L_\mu^l]
&=& 2\sum_{\mu m}D^{l*}_{\mu m}c^\dagger_m[c_i^\dagger c_j,\sum_{n}D^l_{\mu n}c_n]\nn\\
&=&2\sum_{\mu mn}D^{l*}_{\mu m}D^l_{\mu n}c^\dagger_m[c_i^\dagger c_j,c_n]\nn\\
&=&2\sum_{\mu mn}D^{l*}_{\mu m}D^l_{\mu n}(-\delta_{in}c^\dagger_mc_j)\nn\\
&=&-2\sum_m (M_l)_{mi}c_m^{\dagger} c_j,   \eea
and
\bea
\sum_\mu [L_\mu^{l\dagger}L_\mu^{l},c^\dagger_i c_j ]
&=&\sum_{\mu mn}D_{\mu m}^{l*} D_{\mu n}^l [c^\dagger_mc_n,c_i^{\dagger}c_j] \nn \\
&=& \sum_{\mu mn}D_{\mu m}^{l*} D_{\mu n}^l (-\delta_{mj}c_i^\dagger c_n+\delta_{in}c_m^\dagger c_j) \nn \\
&=& \sum_n \left(-(M_l)_{jn} c^\dagger_i c_n +(M_l)_{ni}c_n^\dagger c_j\right).
\eea The corresponding terms in Eq.(\ref{EOM}) sum to $-\{M^T_l,\Delta(t)\}_{ij}$,

Similarly, for the commutators from the gain dissipators, we have
\begin{equation}
\begin{aligned}
2\sum_\mu L_\mu^{g \dagger}[c^\dagger_i c_j,L_\mu^g]
=2\sum_m (M_g)_{mj}c_m c_i^{\dagger},
\end{aligned}
\end{equation}
and
\begin{equation}
\begin{aligned}
\sum_\mu [L_\mu^{g\dagger}L_\mu^{g},c^\dagger_i c_j ]=\sum_n \left( -(M_g)_{nj}c_n c_i^\dagger +(M_g)_{in}c_j c^\dagger_n \right).
\end{aligned}
\end{equation}
Writing $c_n c_i^\dag=\delta_{ni}-c^\dag_i c_n$, we see that the corresponding terms in Eq.(\ref{EOM}) sum to $2(M_g)_{ij}\text{Tr}(\rho)-\{M_g,\Delta(t)\}_{ij} =2(M_g)_{ij} -\{M_g,\Delta(t)\}_{ij}$.  Therefore, all terms at the right hand side of Eq.(\ref{EOM}) sum to that of Eq.(\ref{supp}).

\end{document}